\documentclass[]{aastex631}
\usepackage{amsmath,amsthm,amssymb}
\usepackage{multirow}
\usepackage{hyperref}
\usepackage{xspace}
\usepackage{epstopdf}
\usepackage{graphicx}
\usepackage{hyphenat}

\newcommand{\rmd}{~\mathrm{d}}

\newcommand\superbit{\textsc{SuperBIT}\xspace}
\newcommand\cosmos{\textsc{EL-COSMOS}\xspace}
\newcommand\uoftastro{\affiliation{David A. Dunlap Dept. of Astronomy and Astrophysics, University of Toronto, 50 St. George Street, Toronto, ON, Canada M5S 3H4}}
\newcommand\dunlap{\affiliation{Dunlap Institute for Astronomy and Astrophysics, University of Toronto, 50 St. George Street, Toronto, ON, Canada M5S 3H4}}
\newcommand\uoftphysics{\affiliation{Department of Physics, University of Toronto, 60 St. George Street, Toronto, ON, Canada M5R 2M8}}
\newcommand\utias{\affiliation{University of Toronto Institute for Aerospace Studies (UTIAS), 4925 Dufferin Street, Toronto, ON, Canada M3H 5T6}}
\newcommand\princeton{\affiliation{Department of Physics, Princeton University, Jadwin Hall, Princeton, NJ, USA 08544}}
\newcommand\durhamcfai{\affiliation{Centre for Advanced Instrumentation (CfAI), Durham University, South Road, Durham DH1 3LE, UK}}
\newcommand\stewardobs{\affiliation{Department of Astronomy/Steward Observatory, University of Arizona, 933 North Cherry Avenue, Tucson, AZ, USA 85721}}
\newcommand\jpl{\affiliation{Jet Propulsion Laboratory (JPL), California Institute of Technology, 4800 Oak Grove Drive, Pasadena, CA, USA 91109}}
\newcommand\durhamcfea{\affiliation{Centre for Extragalactic Astronomy, Department of Physics, Durham University, Durham DH1 3LE, UK}}
\newcommand\durhamifcc{\affiliation{Institute for Computational Cosmology, Durham University, South Road, Durham DH1 3LE, UK}}
\newcommand\oslo{\affiliation{Institute of Theoretical Astrophysics, University of Oslo, Blindern, Oslo 0315, Norway}}

\newcommand\washu{\affiliation{Department of Physics, Washington University in St. Louis, 1 Brookings Drive, St. Louis, MO, USA, 63130}}
\newcommand\mcdonnell{\affiliation{McDonnell Center for the Space Sciences, Washington University in St. Louis, 1 Brookings Dr., St. Louis, MO USA 63130}}

\shorttitle{Weak lensing in the blue}
\shortauthors{Shaaban et al.}
\graphicspath{{./}{figures/}}

\begin{document}

\title{Weak lensing in the blue:\\ a counter-intuitive strategy for stratospheric observations}

\correspondingauthor{Mohamed M.\ Shaaban}
\email{m.shaaban@mail.utoronto.ca}

\author[0000-0002-7600-3190]{Mohamed M.\ Shaaban}
\dunlap
\uoftphysics

\author[0000-0002-3937-4662]{Ajay S.\ Gill}
\dunlap{}
\uoftastro{}

\author[0000-0002-9883-7460]{Jacqueline McCleary}
\affiliation{Department of Physics, Northeastern University, 360 Huntington Ave, Boston, MA}

\author[0000-0002-6085-3780]{Richard J.\ Massey}
\durhamifcc
\durhamcfea
\durhamcfai

\author{Steven J.\ Benton}
\princeton{}

\author{Anthony M.\ Brown}
\durhamcfai
\durhamcfea


\author{Christopher J.\ Damaren}
\utias

\author{Tim Eifler}
\stewardobs

\author{Aurelien A.\ Fraisse}
\princeton

\author{Spencer Everett}
\jpl

\author{Mathew N.\ Galloway}
\oslo

\author{Michael Henderson}
\utias

\author{Bradley Holder}
\utias
\dunlap

\author{Eric M.\ Huff}
\jpl

\author{Mathilde Jauzac}
\durhamcfea
\durhamifcc

\author{William C.\ Jones}
\princeton

\author{David Lagattuta}
\durhamcfea

\author{Jason S.-Y.\ Leung}
\uoftastro
\dunlap

\author{Lun Li}
\princeton

\author{Thuy Vy T.\ Luu}
\princeton


\author{Johanna M.\ Nagy}
\washu
\mcdonnell

\author{C.\ Barth Netterfield}
\uoftastro
\dunlap
\uoftphysics

\author[0000-0002-9618-4371]{Susan F.\ Redmond}
\princeton

\author{Jason D.\ Rhodes}
\jpl

\author{Andrew Robertson}
\jpl

\author{J\"urgen Schmoll}
\durhamcfai

\author{Ellen Sirks}
\durhamifcc

\author{Suresh Sivanandam}
\uoftastro
\dunlap




\begin{abstract}

The statistical power of weak lensing measurements is principally driven by the number of high-redshift galaxies whose shapes are resolved. Conventional wisdom and physical intuition suggest this is optimised by deep imaging at long (red or near-IR) wavelengths, to avoid losing redshifted Balmer-break and Lyman-break galaxies. We use the synthetic Emission Line (`EL')-COSMOS catalogue to simulate lensing observations using different filters, from various altitudes. Here were predict the number of exposures to achieve a target $z\gtrsim0.3$ source density, using off-the-shelf and custom filters. Ground-based observations are easily better at red wavelengths, as (more narrowly) are space-based observations. However, we find that \superbit, a diffraction-limited observatory operating in the stratosphere, should instead perform its lensing-quality observations at blue wavelengths.


\end{abstract}



\section{Introduction} \label{sec:intro}


The standard model of cosmology predicts that most of the energy density of the universe is dominated by dark energy and dark matter, both of which possess exotic physical properties that have only been weakly constrained. As a result, constraining the nature of dark energy and dark matter are considered two of the greatest challenges in contemporary science. Weak gravitational lensing is a powerful tool for probing the physics of dark matter and dark energy \citep{Weinberg_2013}. Weak lensing occurs when light radiated from distant galaxies changes path as it passes through a foreground gravitational field. This change of path results in an apparent distortion of the galaxy image. Measuring this change in shape allows us to reconstruct the gravitational field and thus the distribution of (all) mass \citep{mandelbaum_2018_rev}. Mapping the large-scale structure of dark matter can constrain cosmological parameters \citep{2001ApJ...553..545H,2001ApJ...560L.111H}; mapping its distribution around specific entities such as galaxy clusters can also constrain cosmological parameters or test dark matter physics \citep{Review_2010,Harvey_2015,Robertson_2019}. In practice, the original shape of the distorted galaxy is unknown, so weak lensing signals can only be inferred via statistical methods applied to ensembles of galaxies. With the weak lensing signal of a galaxy cluster constrained inside a small patch of sky, the signal-to-noise of cluster mass maps is driven by the spatial {\it density} of galaxies whose shapes can be measured and averaged. 

Conventional wisdom dictates that one can maximize the density of high-redshft galaxies by observing at the red end of the spectrum. The physical intuition behind this is the desire to avoid losing redshifted Balmer-break and Lyman-break galaxies. However, the intuition can be violated by specific observing conditions, environment, and instrumental configurations. One example of an instrument for which this convention is violated is the Super-pressure Balloon-borne Imaging Telescope (\superbit). \superbit is a balloon-borne, diffraction-limited, wide-field, near-ultraviolet to near-infrared observatory designed to exploit the stratosphere's space-like conditions to perform deep lensing observations of galaxy clusters \citep{gill2020optical,romualdez2019robust}. The telescope has a $15\arcmin\times23\arcmin$ field of view that is designed to capture a 4\,Mpc diameter galaxy cluster at $z \ge 0.2$ in a single exposure. \superbit has 4 broadband filters and 2 ultra-broad band filters covering a wavelength range from 300\,nm to 900\,nm. \superbit plans to use either its broad blue band \textit{B} or ultra-broad luminous band $L$ for its deep lensing observations. The purpose of this paper is to provide justification for this seemingly unconventional choice by simulating artificial \superbit observations of the COSMOS survey \citep{cosmos-survey} and comparing the relative performance of the bands. 

\section{Method} \label{sec:general}


Any figure of merit to compare the performance of different filters is likely to track some aspect of image depth as a function of exposure time. For a weak lensing measurement, obtained statistically from ensembles of galaxies, a natural choice is 
the 
density of high redshift galaxies that are both bright enough to be detected and large enough to be resolved. We shall equivalently adopt the observing time required to reach a target threshold in the density of such galaxies. 

To compute this figure of merit, we simulate observations in a range of instrumental configurations and observing environments. The general procedure is as follows. First we fix instrument-specific parameters for the telescope optics and science detector (\autoref{subsec:instspec}). Then we specify the observing environment and determine the flux from sky background that reaches the detector in all of the instrument's bands (\autoref{subsec:enviroment}). Finally we model the flux reaching the detector from distant galaxies (\autoref{subsec:cat}), using galaxies' spectral energy distributions (SEDs) from \citep{Saito_2020}'s synthetic Emission Line COSMOS (\cosmos) catalogue and morphologies from \cite{cosmos_morph}'s COSMOS-Galsim catalogue. We apply cuts similar to those in a real cluster lensing analysis, requiring
\begin{itemize}
    \item galaxy redshift $z>0.3$
    \item galaxy flux signal-to-noise ratio SNR$\geq 5$ inside a fixed $3\times3$\,pixel aperture
    \item galaxy half-light radius $r_e$ greater than the Full Width at Half Maximum (FWHM) of the instrument's Point Spread Function (PSF).
\end{itemize}
Our figure of merit is calculated from the density of galaxies remaining after these cuts, as a function of exposure time.

\subsection{Instrument Specifications}
\label{subsec:instspec}
In this analysis we consider one set of system optics, with two different configurations of detector throughput and optical transmission. Our optics model matches the \superbit telescope \citep{Romualdez_2020}: a diffraction-limited 500\,mm diameter telescope with 38\% of the collecting area obscured by a secondary mirror, and a focal plane Gaussian jitter with $\sigma= 0.05\arcmin\arcmin$. The first configuration of detector throughput (\superbit configuration) comprises the nominal \superbit science detector: the Sony IMX455 CMOS camera, with dark current of 0.002\,electrons/s/pixel, read noise of 1.5\,electrons/pixel, and sensitivity in the 300-900\,nm range measured and presented in \cite{gill2022}. The second configuration (flat configuration) is a fictional configuration with noise specifications identical to the \superbit configuration, but with a total transmission multiplied by quantum efficiency of 0.7\,$e^-$/photon, irrespective of wavelength. This is to enable us to quantify the extent to which our results are dependent on the integrated QE curve shape. Both detector configurations adopt a plate scale of 0.141$\arcmin\arcmin$ per pixel. For both configurations we model the system's Point Spread Function (PSF) as the convolution of a diffraction limited airy disk and a Gaussian kernel representing the system jitter. A summary of the relevant instrument specifications is presented in \autoref{tab:cameras}.

\begin{figure}
    \centering
    \includegraphics[width= 0.7\textwidth]{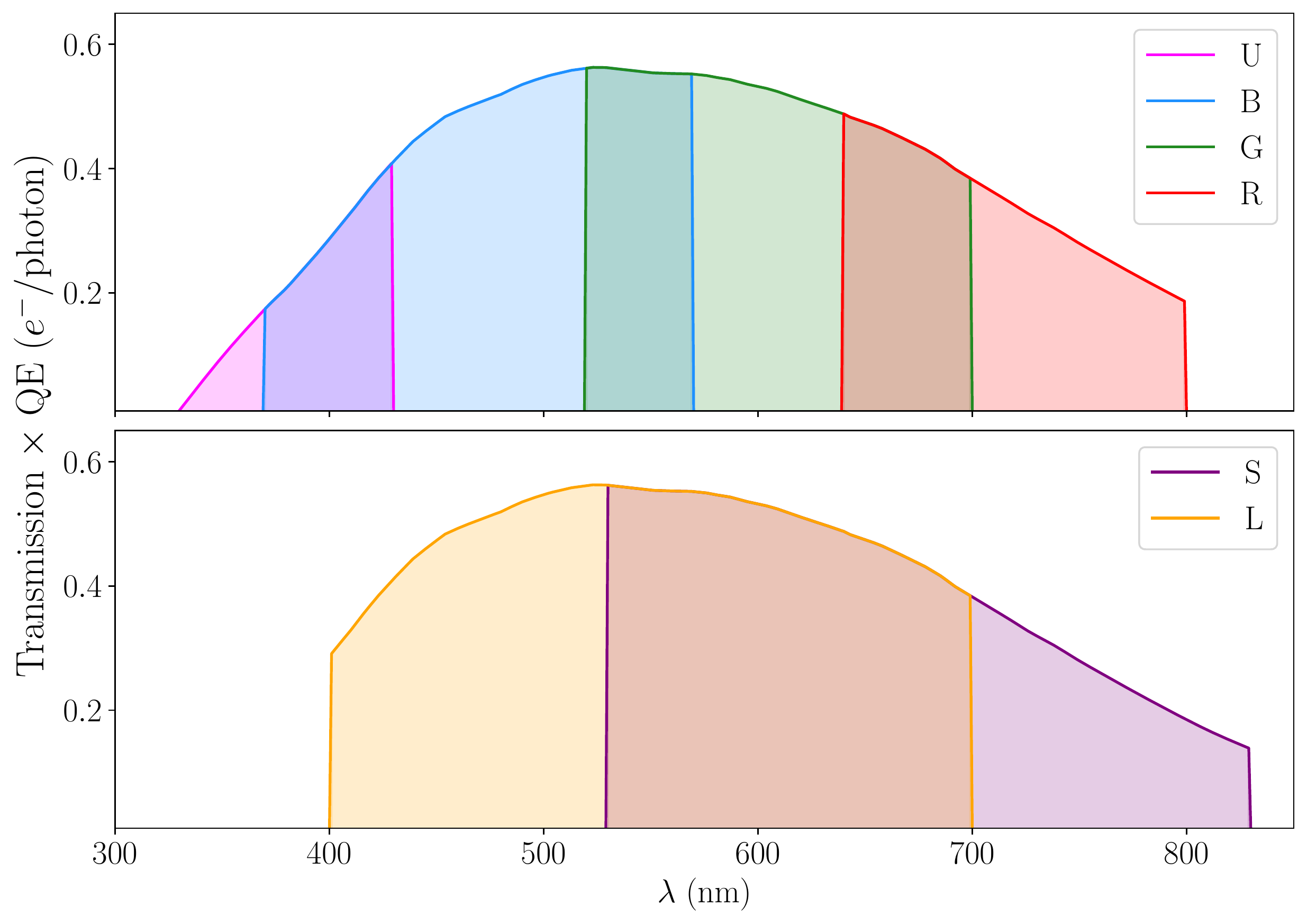}
    \caption{The overall \superbit configuration bandpass which is constructed by taking the product of the throughput of all of \superbit's optical elements, filter transmission, and science camera quantum efficiency. The top panel shows the bandpass for the photometric bands \textit{U}(330-430 nm), \textit{B}(370-570 nm), \textit{G}(520-700 nm), and \textit{R}(640-800 nm). The bottom panel shows the bandpass for the broad bands \textit{L}(400-700 nm) and \textit{S}(530-830 nm). The flat configuration bandpass would be simple a series of horizontal lines at 0.5. }
    \label{fig:filters}
\end{figure}

The final instrument specification of relevance to this analysis is the choice of photometric bandpasses. We perform this analysis in 6 bands. Four are \superbit's nominal photometric bands: \textit{U}(330--430\,nm), \textit{B}(370--570\,nm), \textit{G}(520--700\,nm), and \textit{R}(640--800\,nm). The remaining two are broad bands, one of which is a near-red band that was the originally proposed deep lensing observation band: \textit{S}(530--830\,nm) that has been scrapped as a result of this analysis. The other is the near-blue band currently considered the deep lensing observation band \textit{L}(400--700\,nm). This band was determined using a band optimization tool that will be publicly released in an upcoming publication. The fully integrated bandpass curves for the \superbit configuration are presented in \autoref{fig:filters}. Note that at the time of this analysis, the \superbit science filters were still in the manufacturing process so only the median in-band throughput was known. As a result, the filters are taken to be boxcar with amplitude equal to the median throughput as provided by the vendor ($\geq$0.9) and an out-of-band transmission of 0.

\begin{table}[h!]
\centering
\begin{tabular}{|c|c|}
\hline
Aperture                &  500 mm ($f$/\# = 11)\\ \hline
Obscuring Fraction                  & 0.38\\ \hline
Collecting Area ($A$)                  & 0.121 m$^2$\\ \hline
Focal Plane Jitter (1$\sigma$)    &      0.05$\arcmin\arcmin$   \\ \hline
Bandwidth                 & 300-900 nm (See \autoref{fig:filters}) \\ \hline
Plate Scale                 & 0.141$\arcmin\arcmin$/pixel \\ \hline
Dark Current at -20 C$^\circ$ ($N_d$)                 & 0.002 $e^-/s$/pixel \\ \hline
Read Noise ($N_r$)                & 1.5 $e^-$ \\ \hline

\end{tabular}
\caption{Summary of \superbit's relevant specifications.\label{tab:cameras}}

\end{table}

\subsection{Observing Environment}
\label{subsec:enviroment}
The next step in the analysis is to model the observing environment. Specifically, we need to model the sky background level and atmospheric transmission for both instrument configurations in three distinct environments: the ground, the stratosphere, and low-earth orbit. While \superbit will only collect data in the stratosphere, we include the other more familiar environments to serve as a reference for the reader. 

For the model of sky background in the stratosphere, we use the measured average values from \cite{gill2020optical}. Specifically, we take the flux values at the pivot wavelength $\lambda_p$ of each band in which the sky background was measured, defined as
\begin{equation}
    \lambda_p^2 = \frac{\int (e_\lambda \lambda) \rmd\lambda}{\int (e_\lambda/ \lambda) \rmd\lambda}
    \label{eq:pivot}
\end{equation}

where $e_\lambda$ is the product of the total in-band transmission and quantum efficiency as a function of wavelength, in units of $e^-/$photon. We then linearly interpolate the flux as a function of wavelength between the data points to generate an approximate model of the background SED in the stratosphere. While this method has its limitations, it suffices for this analysis as the bands of interest are similar in bandwidth and pivot wavelength to the bands in which the measurements were performed. For the sky background model in low earth orbit, we use the high sky background data publicly available in the WFC3 manual for the Hubble Space Telescope (HST) \citep{wfc3_handbook}. Finally, for the model of sky background from the ground, we use the publicly available optical background data for the Gemini Observatory\footnote{The Gemini Observatory background data used can be found at \url{http://www.gemini.edu/observing/telescopes-and-sites/sites}}. 

\begin{figure}[h!]
    \centering
    \includegraphics[width= 0.7\textwidth]{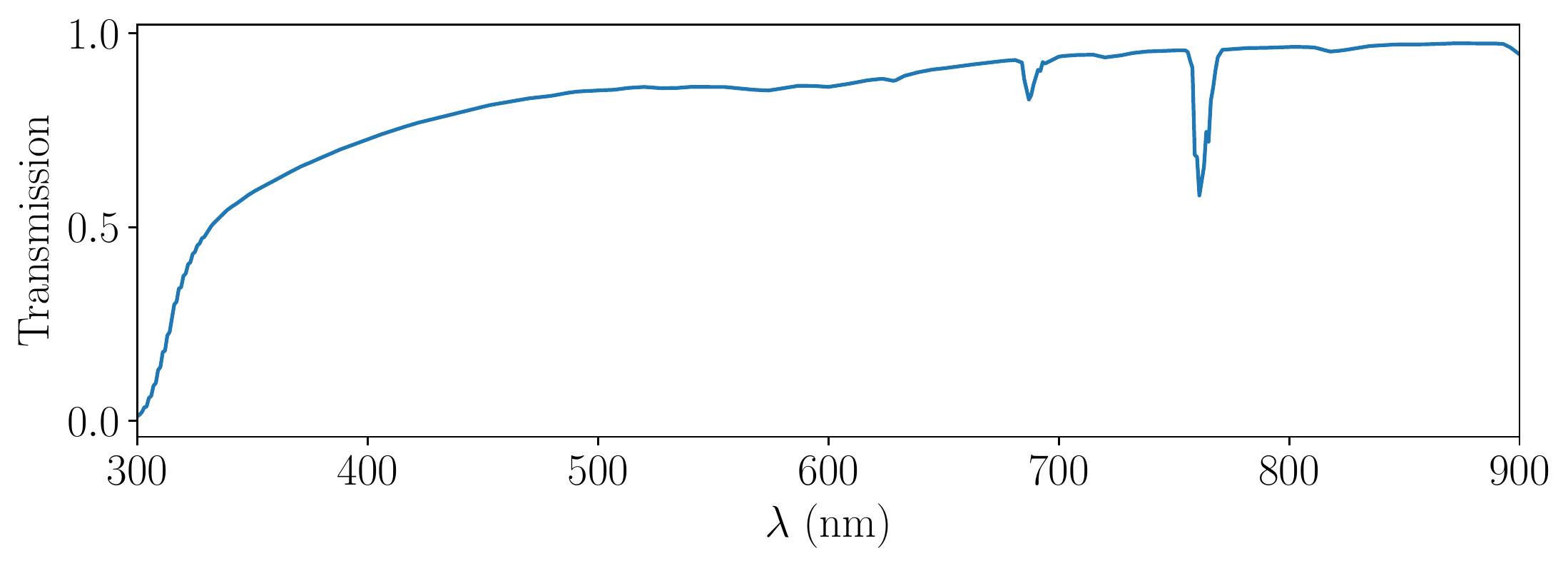}
    \caption{Atmospheric transmission at an altitude of 5\,km above sea level in desert conditions at mid-latitude winter, as calculated by the MODTRAN webapp.}
    \label{fig:btrans}
\end{figure}

The atmospheric transmission in the bands of interest is taken to be order unity in the stratosphere and low-earth orbit. For the ground, we calculated the atmospheric transmission using the MODTRAN webapp \footnote{The MODTRAN webapp can be found here \url{http://modtran.spectral.com/modtran_home}} for an altitude of 5\,km (\autoref{fig:btrans}).

\begin{figure}[h!]
    \centering
    \includegraphics[width= 0.7\textwidth]{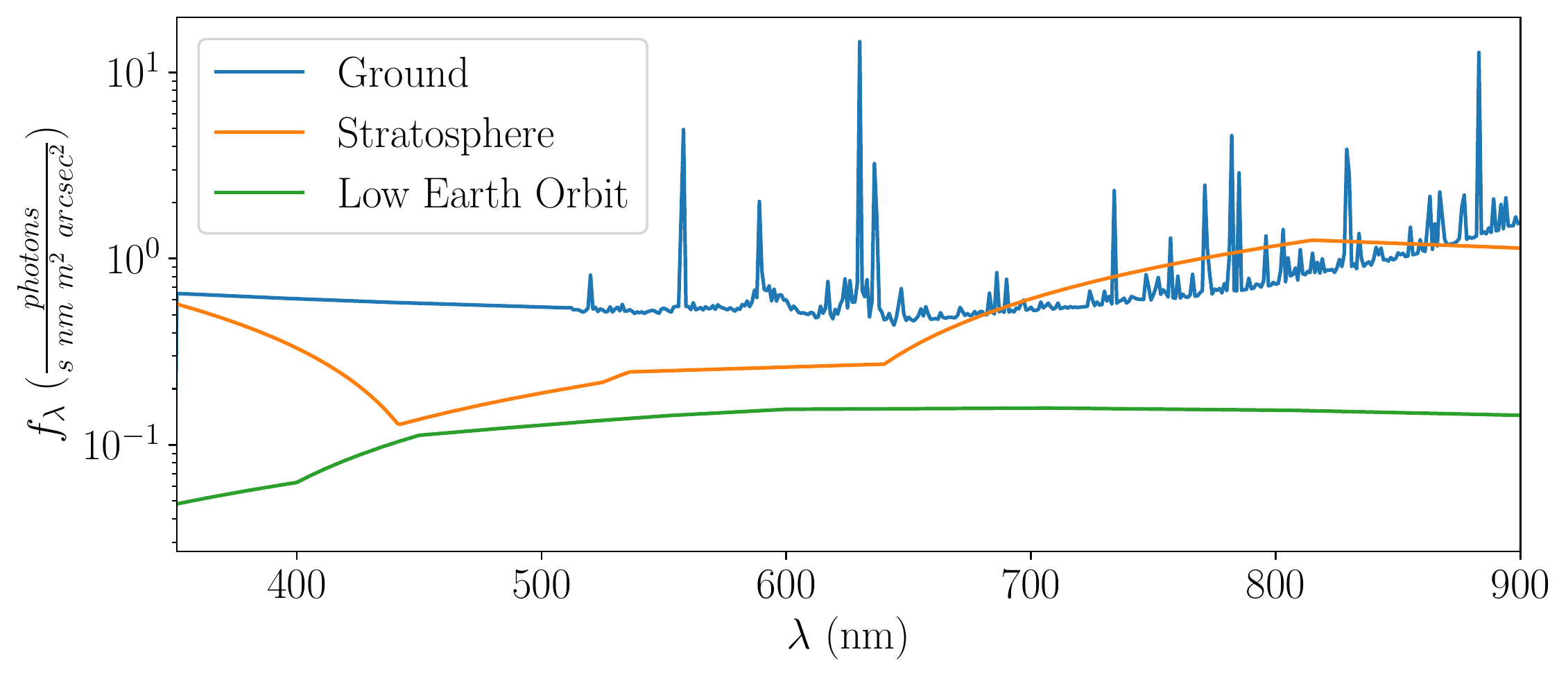}
    \caption{The background flux density in units of $\frac{\text{photons}}{s\ nm\ m^2\ \text{arcsec}^2}$ at the aperture for the three environments considered.}
    \label{fig:backgrounds}
\end{figure}

All three background models are presented in \autoref{fig:backgrounds}. Note that the ground model is an empirical one and therefore the atmospheric transmission is already folded in. We only need to use our modelled transmission curve for simulating the observations of astronomical sources from the ground.
    
\begin{figure}[h!]
    \centering
    \includegraphics[width= 0.7\textwidth]{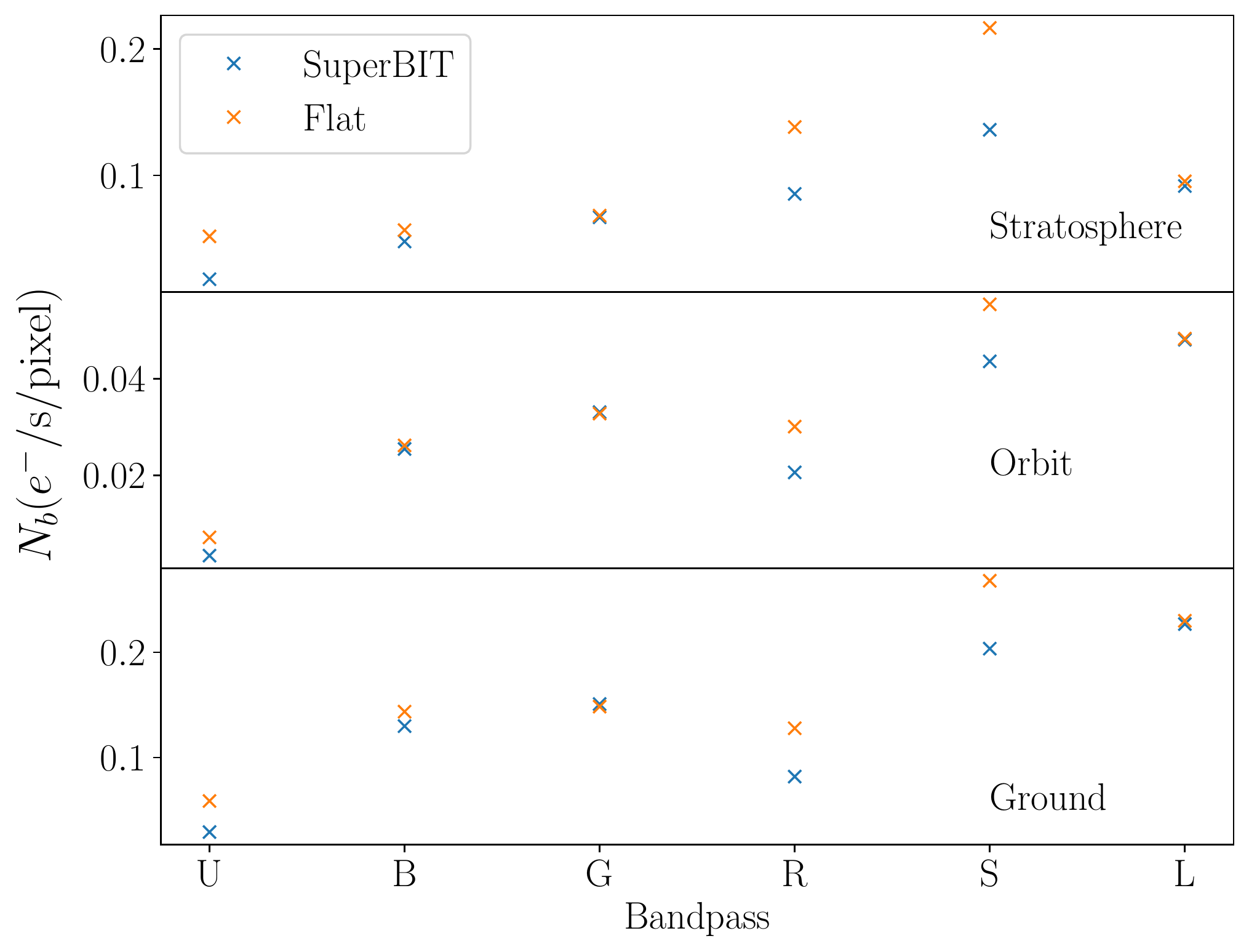}
    \caption{The modelled number of electrons per second per pixel, generated on a detector by the sky background, for the \superbit (blue) and flat (orange) instrument configurations, in all three observing environments.}
    \label{fig:bge}
\end{figure}

Once the environments are modeled, we calculate the total number of electrons per second, $N_b$, generated in each pixel of the detector by the sky background, as
\begin{equation}
    \label{eq:source}
    N_b = AP^2 \int f_{\lambda}e_{\lambda} \rmd\lambda ~,
\end{equation}
where $A$ is the total collecting area, $P$ is the detector's pixel scale in arcseconds/pixel, $f_\lambda$ is the source flux density 
as a function of wavelength, and $e_\lambda$ is the product of the total in band transmission and quantum efficiency as a function of wavelength in units of $e^-/$photon. The calculated background electron flux for all configurations is presented in \autoref{fig:bge}. We note that the background flux per pixel for \superbit in the stratosphere is significantly higher in $S$ than $L$ despite the comparable bandwidth. This long-wavelength sky background plays a major role in \superbit's differential performance at red and blue wavelengths, as we will show in \autoref{sec:results}.   

\subsection{Source Galaxy Population}
\label{subsec:cat}

\subsubsection{Which galaxies are bright enough to be detected?}

\begin{figure}[h!]
    \centering
    \includegraphics[width= 0.7\textwidth]{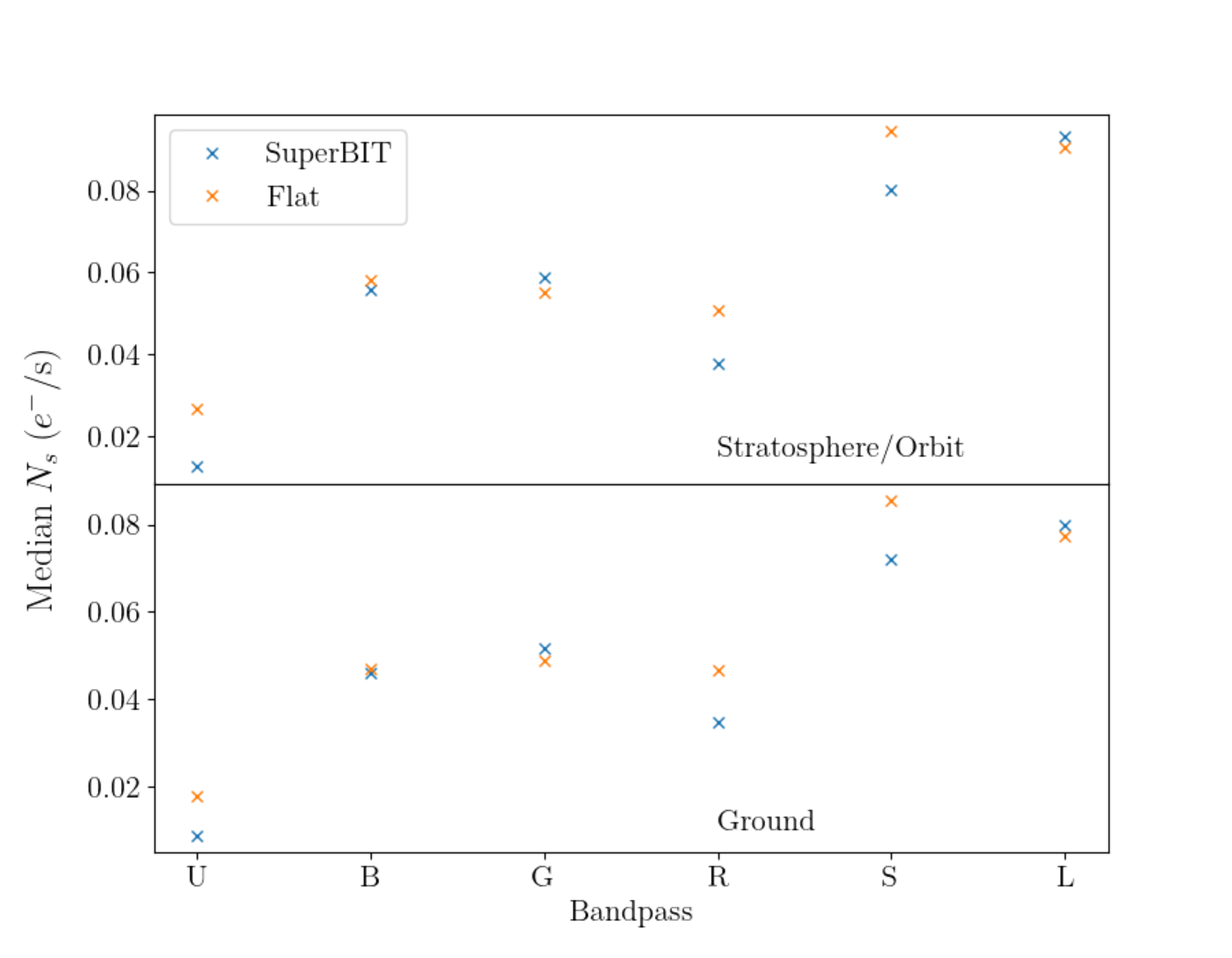}
    \caption{The median number of electrons per second from galaxies in the \cosmos catalogue for the \superbit (blue) and flat (orange) configurations for all three observing environments. Note that due to identical atmospheric transmission curves, the values in orbit and stratosphere are identical.}
    \label{fig:src_e}
\end{figure}

The publicly available \cosmos catalogue \citep{Saito_2020} provides spectral energy distribution (SED) models of more than 500,000 galaxies from the COSMOS2015 catalogue \citep{Laigel_2016}. For each galaxy in the central region\footnote{The \cosmos catalogue is divided into 9 area coverage zones for ease of data management. We use the central zone (zone 5), which is already $\sim$1.9~times larger than the field of view of \superbit. Parts of the analysis were repeated on the other zones with no qualitative change in our results, indicating no zone-selection bias.} of \cosmos, we convert the provided SEDs from energy units to photon units, then calculate the total number of electrons per second, $N_s$ that will be generated on the detector
\begin{equation}
    \label{eq:source}
    N_s = A \int f_{\lambda}T_\lambda e_{\lambda} \rmd\lambda ~,
\end{equation}
where $A$ is the total collecting area, $f_\lambda$ is the source flux density as a function of wavelength, $T_\lambda$ is the atmospheric transmission as a function of wavelength and $e_\lambda$ is the product of the total in-band transmission and quantum efficiency as a function of wavelength, in units of $e^-/$photon. We shall distribute the photons between pixels in section~\ref{subsec:morphology}. As intuition suggested, for the flat instrument configuration, the median source flux is higher in the redder \textit{S} band than the bluer \textit{L} band for all three observing environments (see \autoref{fig:src_e}). However, once the wavelength-dependence of \superbit's quantum efficiency is taken into account, the median on-detector flux is higher in \textit{L}. As will be shown in \autoref{sec:results}, this will further separate  the performance of \superbit between red and blue wavelengths.

To predict the detection SNR of each galaxy in a $3\times3$~pixel aperture\footnote{To check that a 3$\times$3 pixel aperture is a reasonable choice, note that galaxies' median half-light radii are within 0.190$\arcmin\arcmin$ and 0.220$\arcmin\arcmin$ for all relevant redshifts, the \superbit PSF ranges from 0.200$\arcmin\arcmin$-0.400$\arcmin\arcmin$, and that the \superbit detector pixel scale is 0.141$\arcmin\arcmin$.}, we calculate
\begin{equation}
    \label{eq:snr}
    \text{SNR} = \frac{n t \mathcal{F}_p N_s}{\sqrt{n t (\mathcal{F}_pN_s+pN_b+pN_d)+n N_r^2}} ~,
\end{equation}
where $n$ is the total number of stacked exposures, $t$ is the duration of a single exposure in seconds, $\mathcal{F}_p$ is the fraction of the source intensity that lands inside $p=3\times3=9$ pixels on the focal plane, $N_s$ is the total number of electrons per second from the observed source, $N_b$ is the number of electrons per second per pixel from the sky background, $N_d$ is the dark current in electrons per pixel, and $N_r$ is the read noise in electrons per pixel. We determined \{$N_s,N_b,N_d,N_r$\} in previous sections, and the duration of each exposure is dictated by \superbit's pointing and tracking specifications to be $t=300$s \citep[see][]{Romualdez_2020}. The number of such exposures, $n$, will be our independent variable. Our figure of merit will be the value of $n$ that guarantees a density of galaxies with SNR$\geq$5 above a threshold (with low $n$ being better).

\subsubsection{Assumptions about galaxy morphologies and sizes}
\label{subsec:morphology}

\begin{figure}[h!]
    \centering
    \includegraphics[width=0.7\textwidth]{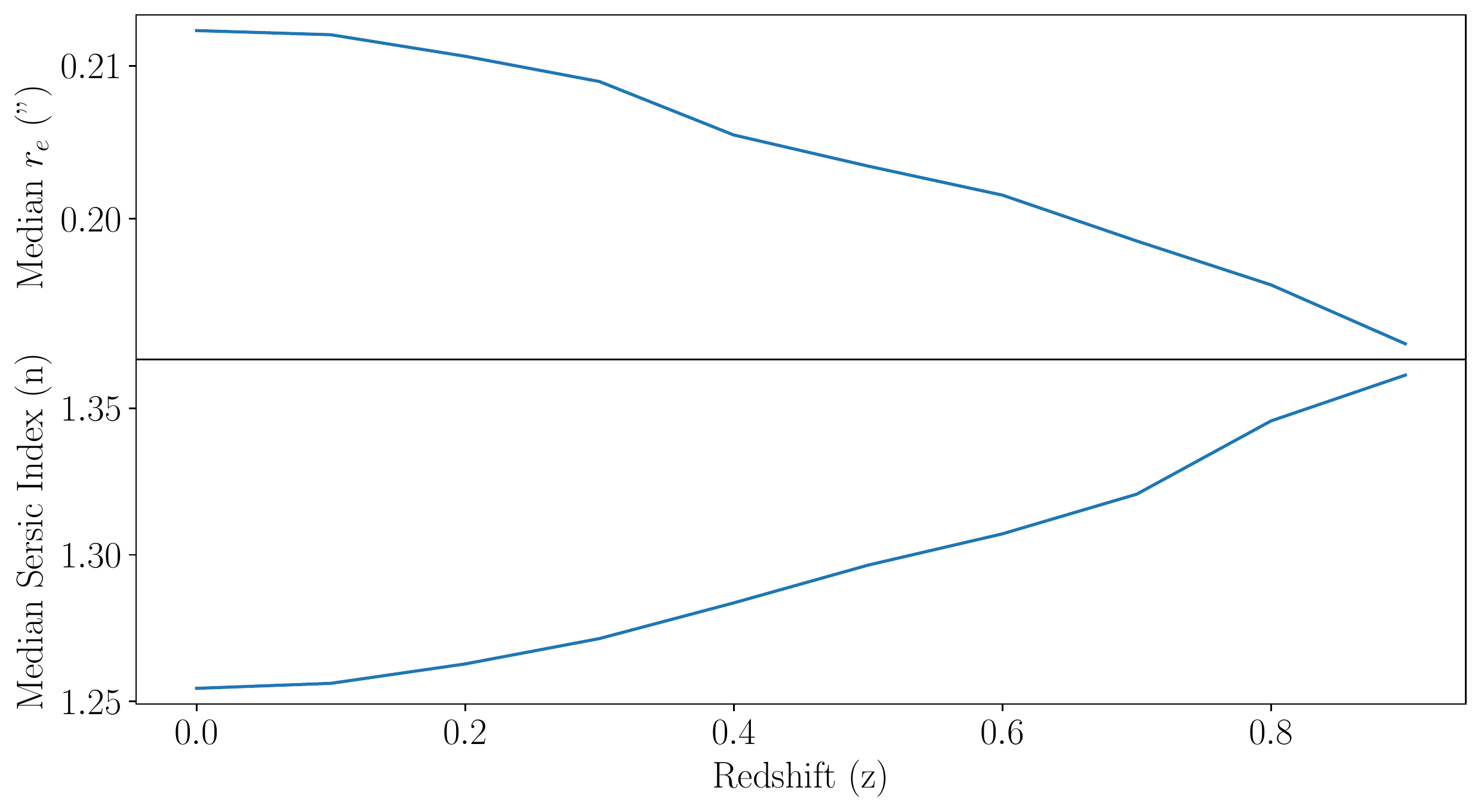}
    \caption{The median S\'ersic half light radius (top panel) and index (bottom panel) of galaxies behind a fixed redshift, as a function of that redshift.}
    \label{fig:Sersic}
\end{figure}

To evaluate the figure of merit from equation~\ref{eq:snr}, we still need to estimate the parameter $\mathcal{F}_p$ for each galaxy. This require knowledge of galaxy morphology that the \cosmos catalog does not provide. We instead exploit fits by \cite{cosmos_morph} of each COSMOS galaxy's surface brightness $I(r)$ to a \cite{Sersic_1963} profile
\begin{equation}
      I(r)= I_{e} \exp\left\{-b_{n}\left[\left({\frac{r}{r_{e}}}\right)^{1/n}-1\right]\right\} ~,
\end{equation}
where $r_e$ is the radius inside which half of the total light of a galaxy is emitted (also known as the half-light radius), $I_e$ is the source intensity at the half-light radius, $n$ is the S\'ersic index parameter, and $b_n$ is a constant determined by the source index. Since our analysis will use all galaxies behind a galaxy cluster at fiducial redshift $z=0.3$, we determine the median values of $r_e$ and $n$ as a function of redshift (see \autoref{fig:Sersic}).


\begin{figure}
    \centering
    \includegraphics[width=0.7\textwidth]{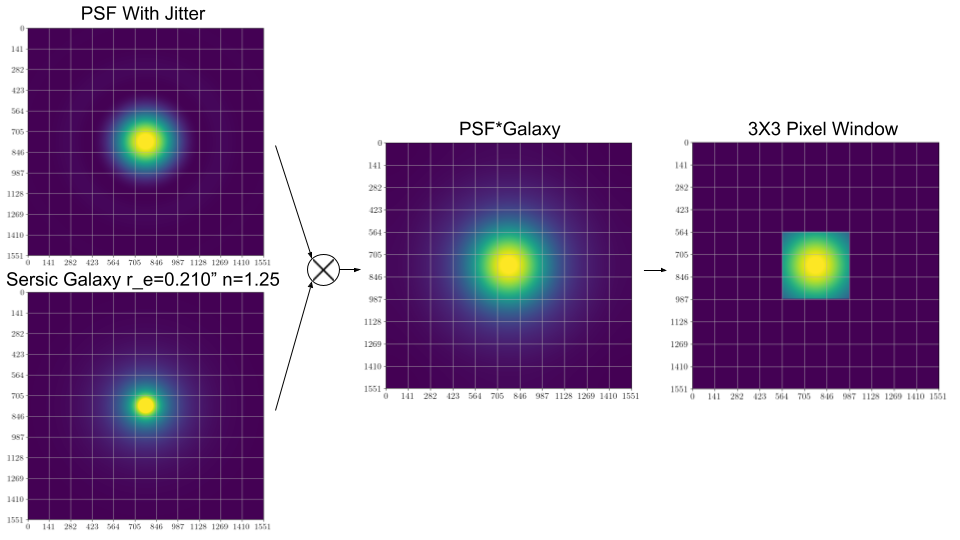}
    \caption{Schematic illustration of the process used to calculate the fraction $\mathcal{F}_9$ of each galaxy's flux that falls within a $3\times3$\, pixel aperture on the focal plane.}
    \label{fig:PSFconv}
\end{figure}

\begin{figure}
    \centering
    \includegraphics[width=0.7\textwidth]{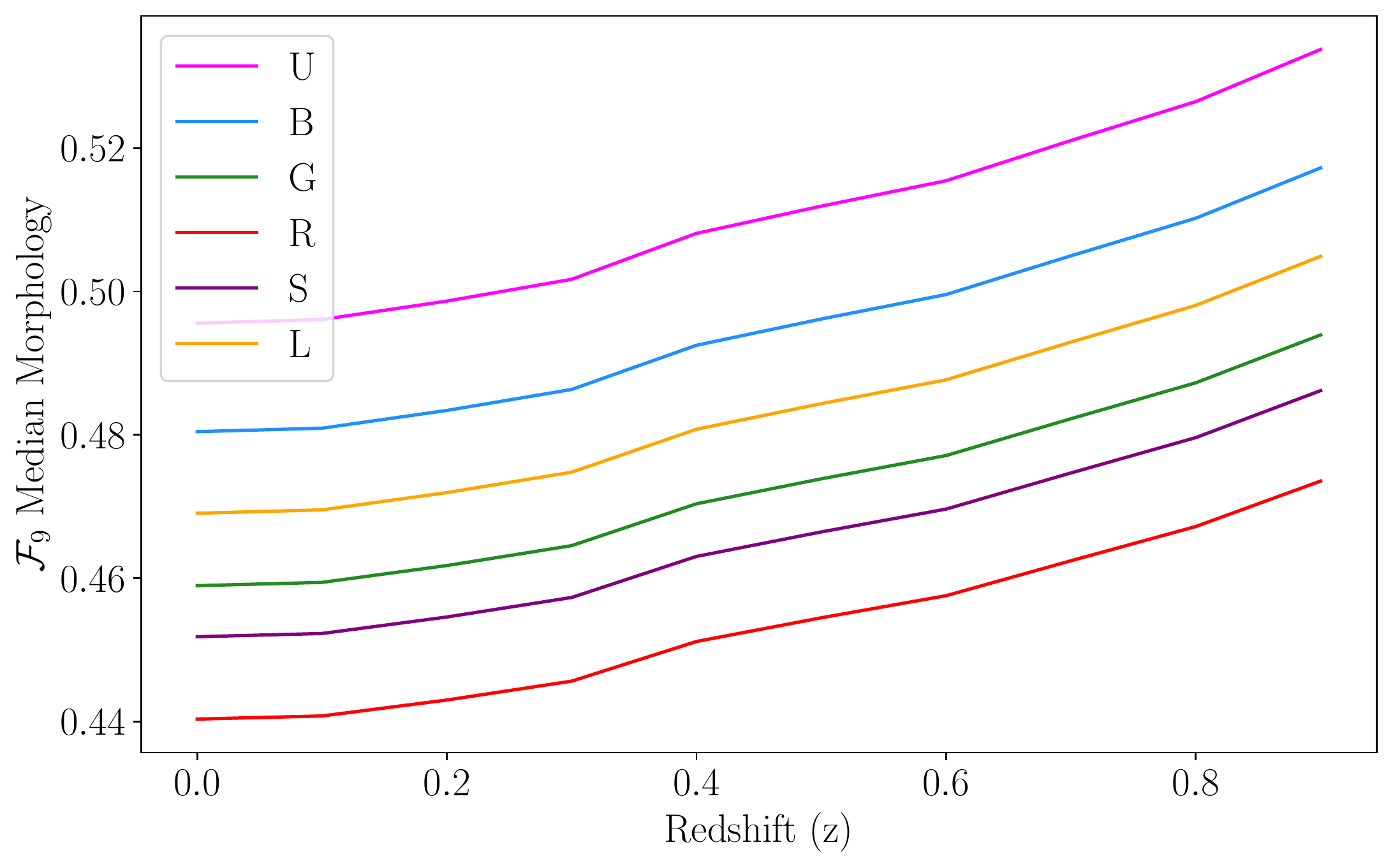}
    \caption{The fraction $\mathcal{F}_9$ of a galaxy's flux that falls within a $3\times3$\, pixel aperture, as a function of cluster redshift for median Sers{\'i}c morphology.}
    \label{fig:fvsz}
\end{figure}

We calculate the median value of $\mathcal{F}_p$ as a function of cluster redshift in all \superbit bands, assuming that galaxies have S\'ersic profiles with parameters drawn from \autoref{fig:Sersic}. We convolve these with a PSF, which is modelled as a diffraction-limited Airy disk centered at the filter's pivot wavelength. In the stratospheric environment, the PSF is also convolved with a Gaussian kernel of width $\sigma=0.05\arcsec$ to represent pointing jitter \citep{Romualdez_2020}. We draw both in an 11$\times$11~pixel window, with each pixel over-sampled by a factor of 141$\times$141 (so each subsampled pixel is conveniently 1\,milliarcsecond on a side). We then convolve the galaxy and PSF models and integrate the fraction of flux inside the central 3$\times$3 pixel aperture (see \autoref{fig:PSFconv}). As one would expect for a fixed aperture, the resulting value of $\mathcal{F}_9$ inversely scales with wavelength (see \autoref{fig:fvsz}), because bluer bands will have a smaller diffraction limited PSF. This effect also contributes to the red-blue performance discrepancy. \footnote{It is important to note that this step induces a non-trivial selection bias in our simulations. In an actual observation the value of p would be dynamically allocated on a source by source basis. However, this step is unavoidable as there currently does not exist public catalogues with sufficient depths to simulate \superbit observations that also provide morphological fits. As a result, we make the most conservative choices to ensure that the selection bias is in favour of larger PSFs (the redder bands) and is therefore dampening the boast in performance gained by the blue bands. The aforementioned choices are: (1) Taking the median source shape of all galaxies behind the cluster rather than in a redshift bin thus putting more weight on smaller sources. (2) Fixing $p=9$ rather than dynamically allocating p on a source by source basis thus favouring PSFs with a size closer to 0.423 \arcsec. }

\begin{figure}
    \centering
    \includegraphics[width=0.7\textwidth]{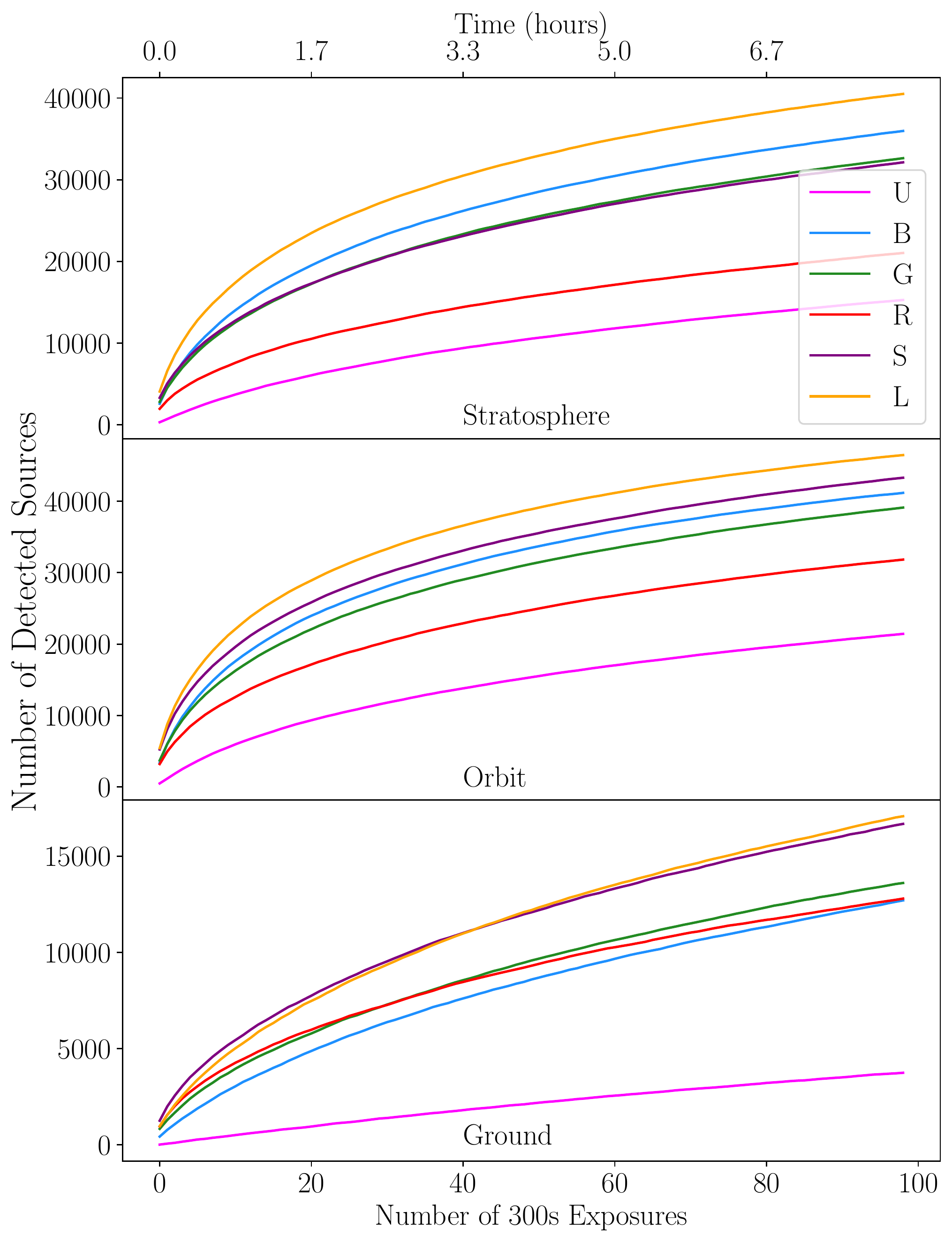}
    \caption{The number of sources with $z>0.3$ that \superbit can expect to detect if it was to observe the \cosmos zone5 as a function of exposure time. The results are presented in all three relevant environments for each one of the bands presented in \autoref{fig:filters}.}
    \label{fig:detections}
\end{figure}

We can now use this information to calculate the number of background sources detected by \superbit when observing a cluster at $z=0.3$. The result is shown in \autoref{fig:detections}.

\subsection{Which galaxies are large enough to be resolved?}

\begin{figure}
    \centering
    \includegraphics[width=0.7\textwidth]{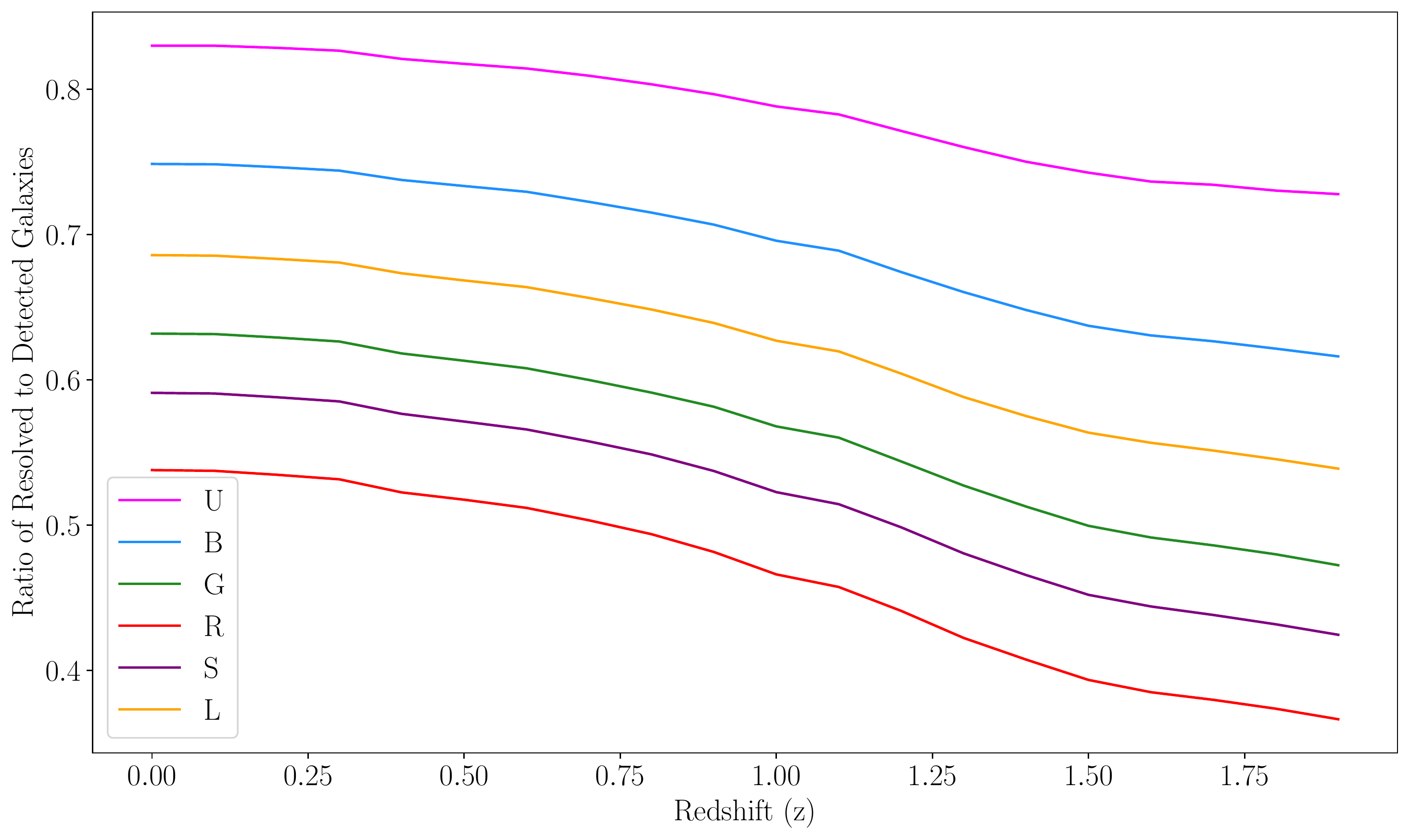}
    \caption{The ratio of resolvable to detectable sources in the \cite{cosmos_morph} catalogue as a function of redshift for each of \superbit's bands.}
    \label{fig:rats}
\end{figure}

Measurements of weak lensing are only possible for galaxies whose shapes are well resolved. Typically, weak lensing analyses impose size cuts of the form $({R_\mathrm{gal}}/{R_\mathrm{PSF}})>r$, where $R_\mathrm{gal}$ is some measure of galaxy size, $R_\mathrm{PSF}$ is some measure of the PSF size, and $r$ is a threshold ratio that depends upon mission systematic errors and knowledge of the PSF \citep{size_cut,Euclid_requirements,des}. We perform a version of this size cut using the galaxy half-light radii from \cite{Mandelbaum_2012b} as $R_\mathrm{gal}$, and the PSF FWHM as $R_\mathrm{PSF}$. Since this analysis is focused on relative performance of bands, the value of $r$ can be arbitrary, so we set it to be unity. This is equivalent to labeling a source as resolved if its half light radius is larger than the FWHM of the in-band PSF. Using these assumptions, we calculate the ratio of resolved to detected sources and present the results in \autoref{fig:rats}. 

There is an important consideration when evaluating the ratios of resolved to detected galaxies in \autoref{fig:rats}. The \cite{cosmos_morph} catalog is shallower than \cosmos, and the relationship between galaxy size and brightness is complex. Therefore, the intrinsic ratio of resolved to detected galaxies would likely differ between the \cite{cosmos_morph} and \cosmos catalogs, had the latter provided galaxy size parameters. This means that the ratio is not directly transferable between catalogues. However, we can reasonably expect that the \textit{relative relationship of the ratio of resolved to detected galaxies between bandpasses} to change very slowly as a function of catalogue depth (this assumption is partially supported by \autoref{fig:rats}).

Therefore, we can transfer the \textit{relative} ratios to \cosmos catalogue by folding them into the curves in \autoref{fig:detections}. It is important to note that this transfer renders the absolute scale of the y-axis meaningless. Instead of a robust physical prediction of what \superbit could expect to observe in flight, the $y$-axis is abstracted to a metric that only possesses comparative information. However, we did calculate the galaxy densities predicted for HST observations, and check that they roughly match the density (as a function of magnitude) in the \cite{leauthaud2017lensing} catalogue. That suggests the normalisation can be meaningful.
%
%

To avoid the misinterpretation of the results as a forecast we perform the following normalization. Since \superbit expects to observe each cluster for order 3 hours (36 exposures), our normalization will be chosen such that 3 hours of \superbit observations in the stratosphere in the \textit{L} band will have a metric value of 1.

\begin{figure}
    \centering
    \includegraphics[width=0.7\textwidth]{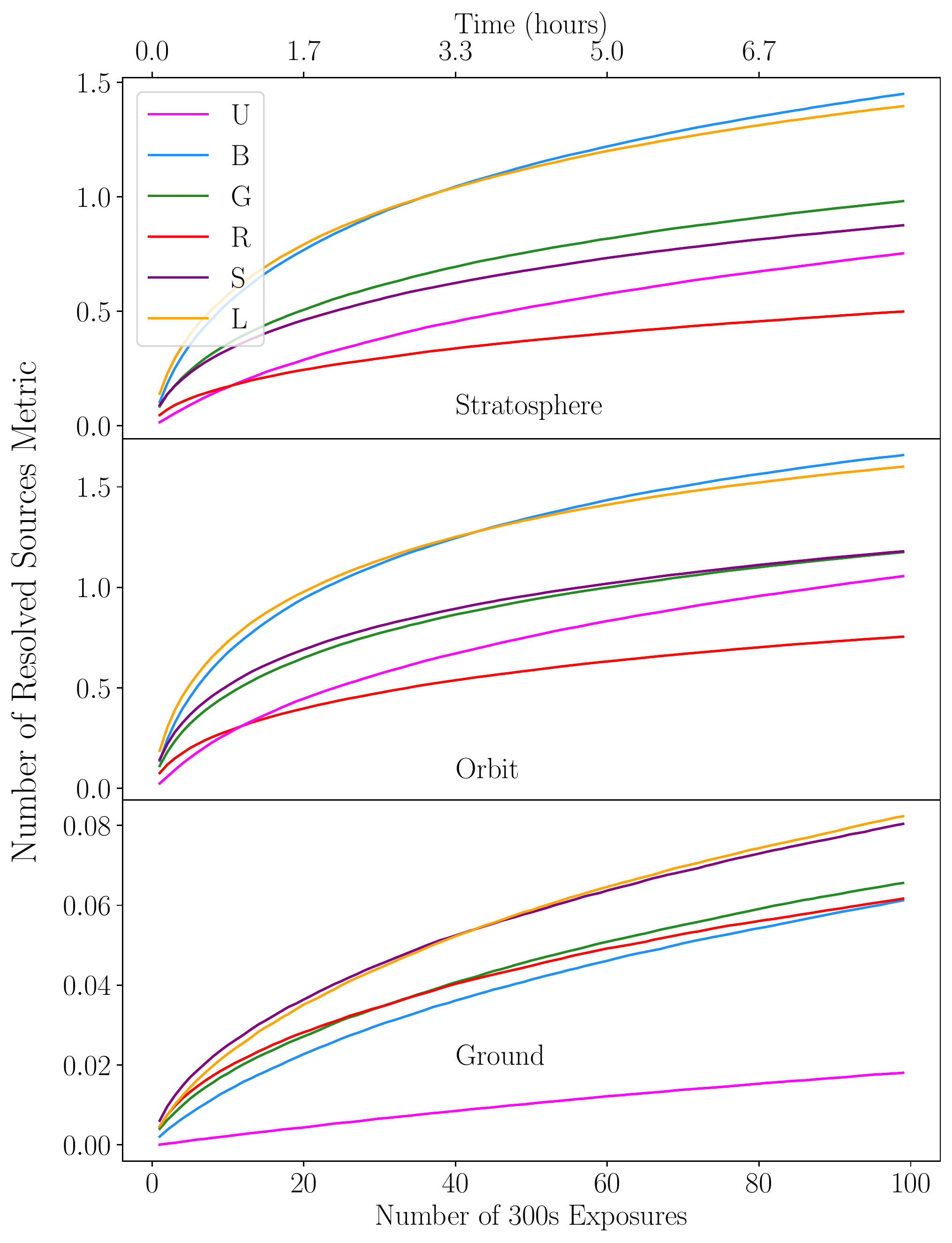}
    \caption{The value of the performance metric for each \superbit band as a function of time, in all three relevant observing environments.}
    \label{fig:resolved}
\end{figure}

\section{Results}
\label{sec:results}
We now have everything we need to evaluate the relative performance of the bands. To do so, we iterate over exposure time by varying the number $n$ of 300s exposures and calculating the SNR of each \cosmos source at each iteration. We simulate an observation of a cluster at $z=0.3$ by discarding all sources with redshift $z\leq0.3$, SNR$\leq$5, and $R_\mathrm{gal}<R_\mathrm{PSF}$, and count the total number of remaining sources in each band as a function of $n$. This simulates the number of detected, resolved sources in each band as a function of exposure time (\autoref{fig:resolved}).

We recover conventional wisdom that lensing measurements from the ground are most efficient in red bands. For observations from the stratosphere however, we find that bluer L (400--700\,nm) and B (370--570\,nm) bands are more efficient, because of the diffraction-limited PSF size compared to the ground, and the near-IR sky background compared to low-Earth orbit. As a result of this analysis, the \superbit team has decided to avoid using the redder \textit{S} band all together and will primarily perform its deep cluster observations in the \textit{B} band, which has simultaneous utility for estimating photometric redshifts.

\subsection{Systematic Errors}

\begin{figure}
    \centering
    \includegraphics[width=0.7\textwidth]{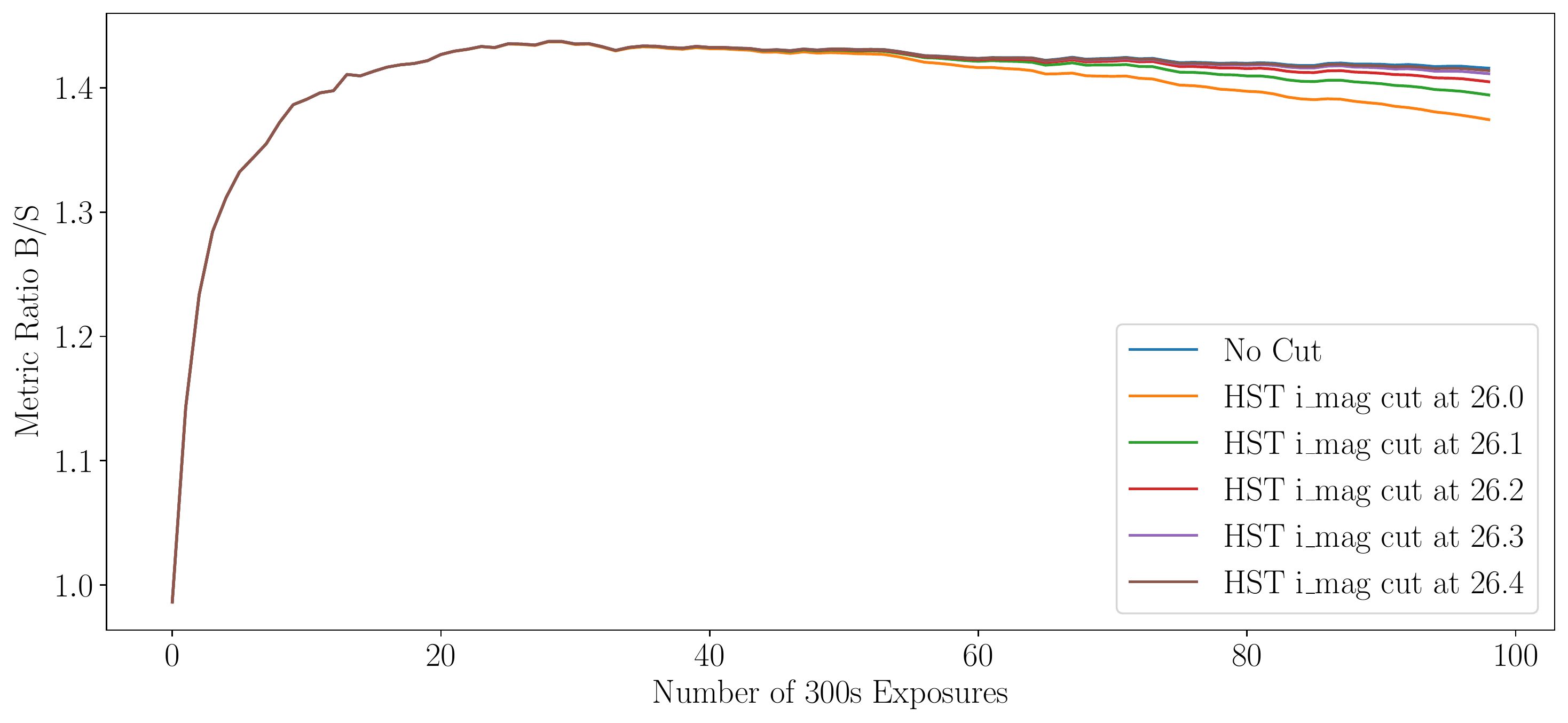}
    \caption{The ratio of the metric value for the $B$ and $S$ bands in the stratospheric environment as a function number of 300 second exposures with varying brightness cuts applied on the catalogue. This figure demonstrates that the bias introduced into our result due to the increasing uncertainty in source redshift is negligible.}
    \label{fig:mag_cuts}
\end{figure}

We recognise that the redshift uncertainty in the \cosmos catalogue increases for dimmer objects. In order to quantify the extent to which this increased uncertainty affects our results, we repeat our analysis on the catalogue with various HSC i-band magnitude cuts. We find that the performance metric ratios between bands is mostly unaffected, as shown in \autoref{fig:mag_cuts}.

Additionally, this analysis does not simulate two systematic errors commonly associated with weak lensing observations of galaxy clusters: source blending and background noise due to intracluster light (ICL). We believe that these effects do not appreciably alter our conclusions for the following reasons: 
\begin{enumerate}
    \item Blending effects become more important for observations with larger PSFs. Inclusion of blending would only strengthen the case for observations in the bluer $B$ band, as its smaller PSF enhances the ability to resolve galaxies. 
    \item Weak lensing measurements are most sensitive to mass distributions projected at 1 Mpc from the cluster center. With \superbit's large field of view, the majority of sources used in WL analysis will lie even farther than this from the cluster's centre. At this radius from the cluster center, the ICL is neglible ($<$ 28 mag arcsec$^{-2}$, see \citealp{2022A&A...666A..54C}), and should not appreciably decrease the observation depth or increase photometric error.
\end{enumerate}




\section{Discussion}
This analysis has involved simulating a variety of observing environments, galaxy sizes, detector configurations, and optical bandpasses. These provide sanity checks in familiar observing environments. They are also physically informative, to test how much each effect contributes to our observed red-blue performance discrepancy. To investigate these effects' relative contributions, we toggle the effects creating a variety of effect-free configurations. The configurations considered are the following:

\begin{itemize}
    \item \textbf{No size cut ($r=0$):} In this configuration, we set the size cutoff ratio $r$ to 0 which is equivalent to treating all detected sources as resolved and therefore bluer bands lose the performance advantage gained from having a smaller PSF.
    \item \textbf{No QE shape (Flat):} In this configuration, the integrated system's quantum efficiency curve is taken to be flat at 0.5 $e^-$/photon eliminating the performance advantage blue bands gain due to \superbit s superior sensitivity in the blue.
    \item \textbf{No background flux ($N_b=0$):} In this configuration, the background flux $N_b$ is set to 0 thus eliminating the performance advantage gained by blue bands due to the bright red background in the stratosphere. 
    \item \textbf{No source smearing ($\mathcal{F}_9=1$):} In this configuration, we enforce that the entire flux from the source lands within the aperture thus eliminating the advantage gained by bluer bands due to lower smearing resulting from smaller PSFs. 
    \item \textbf{Raw flux density:} In this configuration, all the effects mentioned above are turned off (i.e. $r=0$ with a flat quantum efficiency, no background flux, and no image smearing).
\end{itemize}

Each of the above effects is toggled on and off, and the metric value at three hours (36 nominal exposures) is calculated in the stratospheric environment. The results of this procedure are presented in \autoref{tab:rats} and \autoref{tab:rats2}. 

\begin{table}[h!]
\centering
\begin{tabular}{|c|c|c|c|c|c|c|}
\hline
&\superbit&	r = 0&	Flat	& $N_b$ = 0	& $\mathcal{F}_9$ = 1&	Raw \\
\hline

\textbf{U} & 0.4                        & 0.56           & 0.61          & 0.69             & 0.83           & 1.73         \\
\textbf{B} & 0.98                       & 1.59           & 0.96          & 1.69             & 1.59           & 3.43         \\
\textbf{G} & 0.67                       & 1.41           & 0.63          & 1.31             & 1.16           & 3.58         \\
\textbf{R} & 0.32                       & 0.87           & 0.34          & 0.75             & 0.62           & 3.2          \\
\textbf{S} & 0.6                        & 1.4            & 0.57          & 1.34             & 1.03           & 3.8          \\
\textbf{L} & 1                          & 1.86           & 0.97          & 1.8              & 1.57           & 3.73         \\ 
\hline
\end{tabular}
\centering
\caption{The performance metric value after 36 exposures for each \superbit band with each effect independently toggled off in the stratosphere. }
\label{tab:rats}
\end{table}

\begin{table}[]
\centering
\begin{tabular}{|c|c|c|c|c|c|}
\hline
 & r = 0&	Flat	& $N_b$ = 0	& $\mathcal{F}_9$ = 1&	Raw \\
\hline
\textbf{U} & 1.40                    & 1.53                   & 1.73                      & 2.08                    & 4.33                  \\
\textbf{B} & 1.62                    & 0.98                   & 1.72                      & 1.62                    & 3.50                  \\
\textbf{G} & 2.10                    & 0.94                   & 1.96                      & 1.73                    & 5.34                  \\
\textbf{R} & 2.72                    & 1.06                   & 2.34                      & 1.94                    & 10.00                 \\
\textbf{S} & 2.33                    & 0.95                   & 2.23                      & 1.72                    & 6.33                  \\
\textbf{L} & 1.86                    & 0.97                   & 1.80                      & 1.57                    & 3.73                  \\ \hline
\end{tabular}
\caption{This table is simply the result of dividing columns 2-6 in \autoref{tab:rats} with the the corresponding column 1 entry. This highlights how each each effect contributes to the final result. }
\label{tab:rats2}
\end{table}

This result shows that while all of these effects favour bluer bands, the dominant effects resulting in the superior blue performance are the background, and the PSF size difference at the half meter diffraction limit. Surprisingly,\superbit's superior QE in the blue plays only a minor role. 

\begin{figure}
    \centering
    \includegraphics[width=0.7\textwidth]{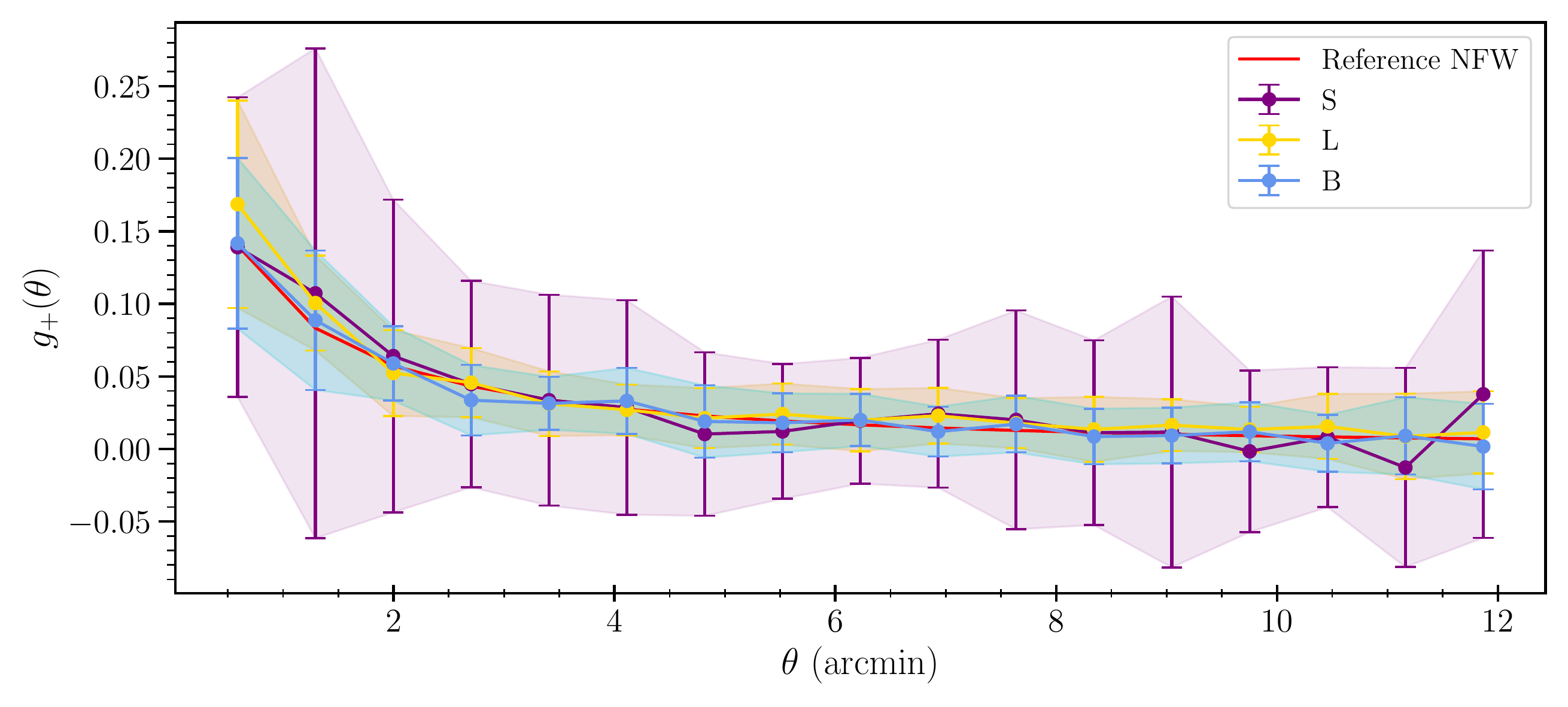}
    \caption{Median tangential shear profiles over 30 simulated observations of a $4.1\times10^{14}$ Msol galaxy cluster at $z=0.45$. Error bars represent the standard deviation of the mean tangential shear in a radial bin. The red line is the theoretical NFW shear input to all simulations. The variance is more than three times larger in \textit{S} than \textit{B} or \textit{L}. Reproduced from the upcoming publication McCleary et al. 2022}
    \label{fig:jmac}
\end{figure}

The \superbit forecasting analysis presented in the upcoming publication McCleary et al. 2022, strongly supports the central thesis of this paper. The analysis found that when weak lensing pipeline was run on simulated \textit{B} images it was able to use 7447 background sources as opposed to the 2578 it was able to use in \textit{S} images. This result is summarised in \autoref{fig:jmac}.

\section{Acknowledgements}

Support for the development of \superbit is provided by NASA through APRA grant NNX16AF65G. 
Launch and operational support for the sequence of test flights from Palestine, Texas are provided by the Columbia Scientific Balloon Facility (CSBF) under contract from NASA's Balloon Program Office (BPO). Launch and operational support for test flights from Timmins, Ontario are provided by the \textit{Centre National d'\'Etudes Spatiales} (CNES) and the \textit{Canadian Space Agency} (CSA).

JR, EH, and SE are supported by JPL, which is run under a contract by Caltech for NASA. Canadian coauthors acknowledge support from the Canadian Institute for Advanced Research (CIFAR) as well as the Natural Science and Engineering Research Council (NSERC). The Dunlap Institute is funded through an endowment established by the David Dunlap family and the University of Toronto. UK coauthors acknowledge funding from the Durham University Astronomy Projects Award, STFC [grant ST/P000541/1], and the Royal Society [grants UF150687 and RGF/EA/180026].



\bibliographystyle{aasjournal}

\bibliography{ref}

\end{document}